\documentclass[prb,twocolumn,superscriptaddress,showpacs]{revtex4}
\usepackage{graphicx}
\usepackage{epsfig}
\usepackage{amssymb,amsmath,amsfonts,hyperref}
\usepackage{wasysym}
\usepackage{latexsym}
\usepackage{color}

\newcommand{\be}{\begin{eqnarray}}
\newcommand{\ee}{\end{eqnarray}}

\begin{document}
\title{Vacancy-induced spin texture in a one dimensional $S=1/2$ Heisenberg
antiferromagnet}
\author{Sambuddha Sanyal}
\affiliation{Department of Theoretical Physics, Tata Institute of Fundamental Research}
\author{Argha Banerjee}
\affiliation{Department of Theoretical Physics, Tata Institute of Fundamental Research}
\author{Kedar Damle}
\affiliation{Department of Theoretical Physics, Tata Institute of Fundamental Research}

\begin{abstract}
We study the effect of a missing spin in a one dimensional $S=1/2$ antiferromagnet with nearest neighbour Heisenberg exchange $J$ and six-spin
coupling $Q=4qJ$ using Quantum Monte-Carlo (QMC) and bosonization techniques.
For $q< q_c \approx 0.04$, the system is
in a quasi-long range ordered power-law
antiferromagnetic phase, which gives way to
a valence-bond solid state that spontaneously
breaks lattice translation symmetry for $q> q_c$. We study the ground
state spin texture $\Phi(r) = \langle G_{\uparrow}|S^z(r)|G_{\uparrow}\rangle$
in the the $S^z_{tot}=1/2$ ground state $|G_{\uparrow}rangle$ of the system with a missing spin, focusing on the alternating part
$N_z(r)$. We find that our QMC results for $N_z$
at $q =q_c$ take on the scaling form expected from bosonization
considerations, but violate scaling
for $q < q_c$. Within the bosonization approach,
such violations of scaling arise from the presence of
a marginally irrelevant sine-Gordon interaction,
whose effects we calculate using renormalization group (RG) improved perturbation theory. Our field-theoretical predictions are found to agree well with
the QMC data for $q < q_c$.

\end{abstract}

\pacs{75.10.Jm 05.30.Jp 71.27.+a}
\vskip2pc

\maketitle

\section{Introduction}
The one-dimensional $S=1/2$ Heisenberg antiferromagnetic spin
chain, with nearest-neighbour exchange couplings $J$ is perhaps
the simplest important model spin system in quantum magnetism. It has not only proved
useful as a theoretical model for the magnetic properties of several
Mott insulating materials\cite{exptchains,exptchains1,exptchains2,exptchains3},
but has also been the subject of many theoretical advances such as Bethe's
original `Bethe Ansatz' solution of this quantum many-body problem and
later field-theoretical treatments that applied bosonization
techniques to map the system to a $1+1$ dimensional bosonic field theory
with a so-called `sine-Gordon' action, made up of a scale invariant free-field
part perturbed by a non-linear cosine interaction.\cite{Affleck_review}
In addition, the renormalization group (RG) analysis of the cosine interaction
that perturbs the scale-invariant free-field action is a paradigmatic example of the treatment
of `marginally irrelevant' interactions in
the neighbourhood of a well-characterized and tractable scale invariant RG fixed point.\cite{Singh_Fisher_Shankar,Affleck_Gepner_Shultz_Ziman,Barzykin_Affleck,Orignac,Diptiman,Eggert}

Such marginally irrelevant interactions 
can give rise to violations of scaling predictions at critical
points due to the presence of logarithmic corrections
that multiply the scaling answer. A
well known example is the $O(N)$ critical point
in four space-time dimensions.\cite{Kenna_ONmodel} In some other cases,
such marginally irrelevant interactions give rise
to {\em additive} {\em corrections} to scaling, which
vanish logarithmically slowly. The one dimensional Heisenberg
chain displays both kinds of effects. For instance, gaps
in the finite size spectra of the spin-half chain are known
to have additive logarithmic corrections that do not affect
the leading behaviour\cite{Affleck_Gepner_Shultz_Ziman},
while the temperature dependence of the NMR relaxation rate $1/T_1$ 
violates scaling expectations due to the presence of an
additional logarithmic factor in its temperature dependence.\cite{Sachdev}

Similar logarithmic violations of scaling, arising from multiplicative logarithmic factors that multiply scaling predictions, have
been argued to exist\cite{Sandvik,Banerjee_Damle_Alet1} in a much less well-understood case of a two dimensional
$S=1/2$ square lattice Heisenberg antiferromagnet on the verge
of a continuous quantum phase transition\cite{Senthil_etal,Sandvik_original}  between the usual Neel ordered
antiferromagnetic ground state and a spontaneously dimerized non-magnetic state
with valence-bond order.
The underlying
critical non-compact CP$^1$ (NCCP$^1$) field theory that has been proposed\cite{Senthil_etal}
as the continuum description of this transition is not as well
understood from a RG standpoint, and
since the numerics themselves are also more challenging, there
have been some differences in the interpretation of these
results.\cite{Sandvik_Kotov_Sushkov,Kaul}

In our own recent work,\cite{Banerjee_Damle_Alet1} we have
used extensive 
numerical computations to establish the presence of
apparently logarithmic scaling violations in the impurity spin texture
induced by a missing-spin defect at such a quantum critical
point when the system has the usual $SU(2)$ symmetry of spin rotations, and
ascribed this effect to the presence of a yet-to-be-identified
marginal operator at the putative NCCP$^1$ critical fixed point. In contrast,
the corresponding spin texture in a system at an analogous critical point with enlarged $SU(3)$ symmetry\cite{Sandvik_SU3} was found to obey scaling predictions without
any logarithmic violations,\cite{Banerjee_Damle_Alet2}, suggesting
that the underlying NCCP$^2$ critical point describing this
$SU(3)$ transition is free of such marginal
operators. However, parallel work of Kaul\cite{Kaul} argues
that such marginal operators would typically
not lead to violations of scaling,
and finds an alternative scenario more likely. In this alternative
scenario, 
both the $SU(2)$ and $SU(3)$ transitions are
described by fixed points with a leading irrelevant operator with
small scaling dimension, and the violations
of scaling arise from the fact that the quantity being studied depends
{\em non-analytically} on this leading irrelevant operator.

Here, we try and understand the origins of such multiplicative
logarithmic corrections to impurity spin textures by using
the one dimensional Heisenberg antiferromagnet as an example. On
the analytical side, we work within the bosonization framework and use
renormalization group (RG) improved perturbation
theory to obtain predictions for the alternating part of the spin texture
in this example. These predictions are compared with Quantum Monte-Carlo (QMC)
results for a one-dimensional chain with nearest neighbour
Heisenberg exchange $J$ and six-spin coupling $Q=4qJ$.
The Hamiltonian for this `$JQ_3$ model' is:
\begin{equation}
H = -J\sum_{i=0}^{N}P_{i,i+1} -Q\sum_iP_{i,i+1} P_{i+2,i+3}P_{i+4,i+5}
\label{JQ3model}
\end{equation}
where $P_{ij} \equiv (\frac{1}{4} - \vec{S}_i \cdot \vec{S}_{j})$
is the projector to the singlet state of the two spin-half variables
at sites $i$ and $j$, both $J$ and $Q$ are assumed positive, and we
impose periodic boundary conditions by placing
the system on a ring so that site $N+1+k$ is identified
with site $k$ (the total number of spins $N+1$ is taken even).

From our QMC results, obtained using the singlet sector valence-bond
projection method\cite{Sandvik_Evertz}, we find that the $Q$ term drives a transition to a valence-bond solid phase at $q_c \approx 0.04$, so
that the system is power-law Neel ordered for $q <q_c$, and VBS ordered
for $q>q_c$. Unlike the more well-studied case
in which such a transition is driven by next-nearest neighbour Heisenberg 
antiferromagnetic exchange couplings, the present $JQ_3$ model does
not have a sign problem in standard non-zero temperature QMC calculations
(as well as in the ground state projector QMC approach), and can therefore be studied at larger length scales and greater precision.

In order to explore the effects of vacancy defects,
we remove the spin at site $0$ and delete all interactions
that involve this spin from our Hamiltonian. Since $N$ is odd,
the ground state of the chain with a missing spin is a doublet with 
$S_{tot} = 1/2$. We focus on $|G_{\uparrow}\rangle$, the $S^z_{tot}=1/2$ component of this doublet, and compute the spin texture $\Phi(r) = \langle S^z(r)\rangle_{\uparrow}$ in this ground state for various values of $q$.
This is done using a recently developed modification\cite{Banerjee_Damle}
of the singlet-sector projector Quantum Monte
Carlo (QMC) technique.\cite{Sandvik_Evertz}. This spin texture
can be decomposed as $\Phi(r) = \Phi_u(r) + (-1)^{r/a}N_z(r)$, where
alternating part $N_z(r)$ and a uniform
part $\Phi_u(r)$ are obtained from our numerical data by
a suitable coarse-graining procedure.
\begin{figure}
{\includegraphics[width=\columnwidth]{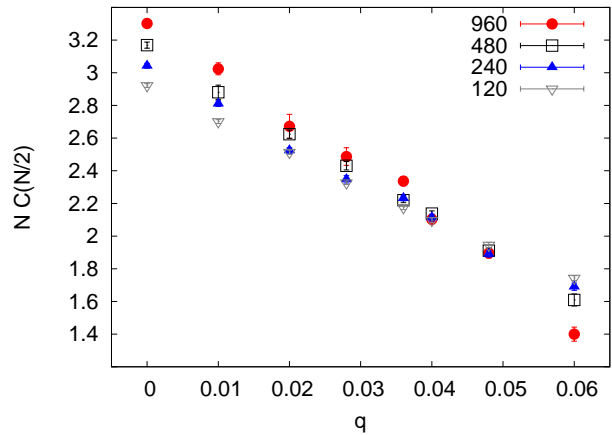}}
\caption{Spin-spin correlation function at distance $N/2$ $C(N/2) = \langle \vec{S}_i \cdot \vec{S}_{i+N} \rangle$ in the ground state of
a periodic chain with $N$ spins, multiplied by $N$ and plotted
against $q$ to serve as diagnostic of the quantum phase transition
from power-law Neel order to valence-bond solid order, as discussed
in the text.}
\label{crossing}
\end{figure}

These numerical results for $N_z(r)$ are compared
to field theoretical calculations within the bosonization framework, keeping careful track of the effects of the marginal
cosine interaction term using one loop RG improved perturbation theory.
Our basic conclusion is that this marginal cosine interaction does
indeed lead to logarithmic violations of scaling by introducing
logarithmic corrections that multiply the scaling predictions for
$N_z$ in the power-law Neel phase. Comparing
these analytical predictions with our numerical
results for $q <q_c$, we find good agreement with the data, with
the strength of the log corrections being larger for $q$ further away
from the critical point, and vanishing for $q=q_c$, as predicted
by the bosonization approach.
\begin{figure}
{\includegraphics[width=\columnwidth]{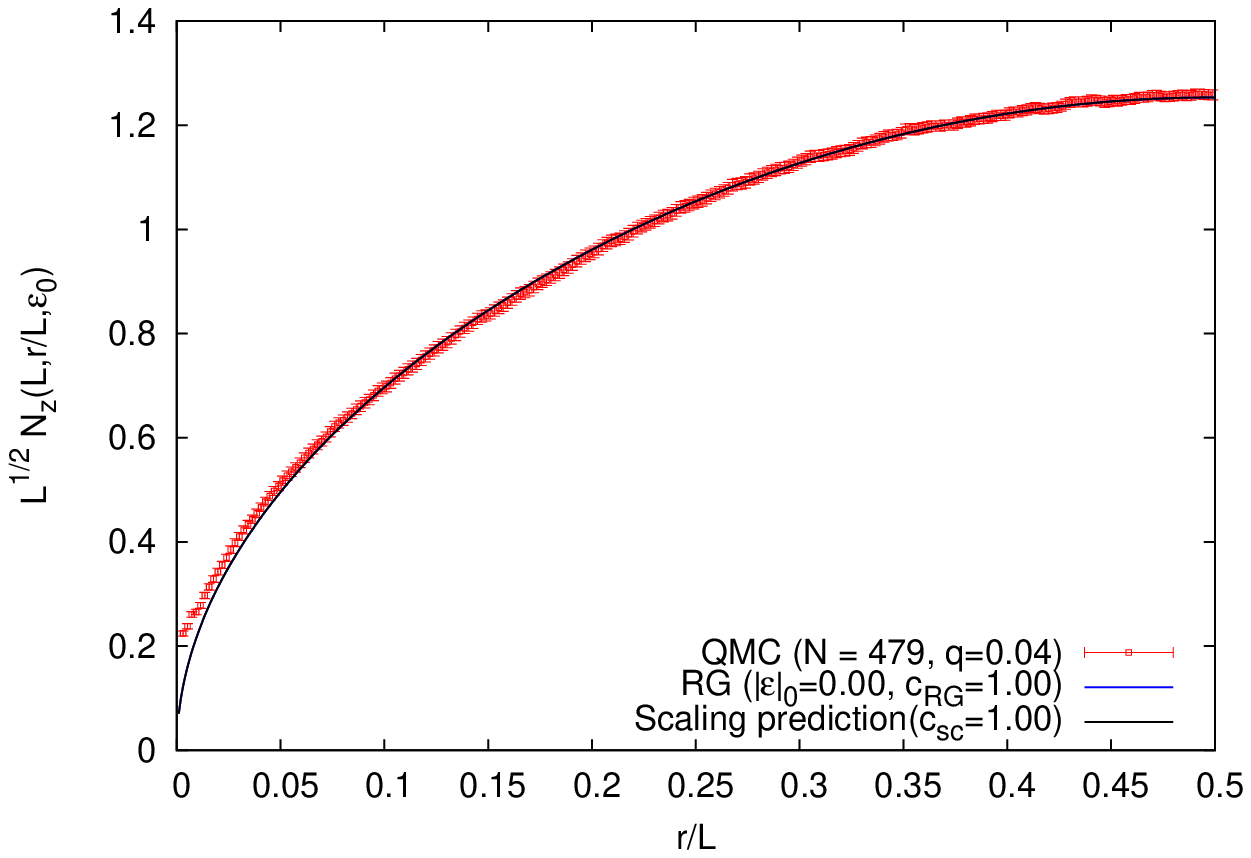}}
{\includegraphics[width=\columnwidth]{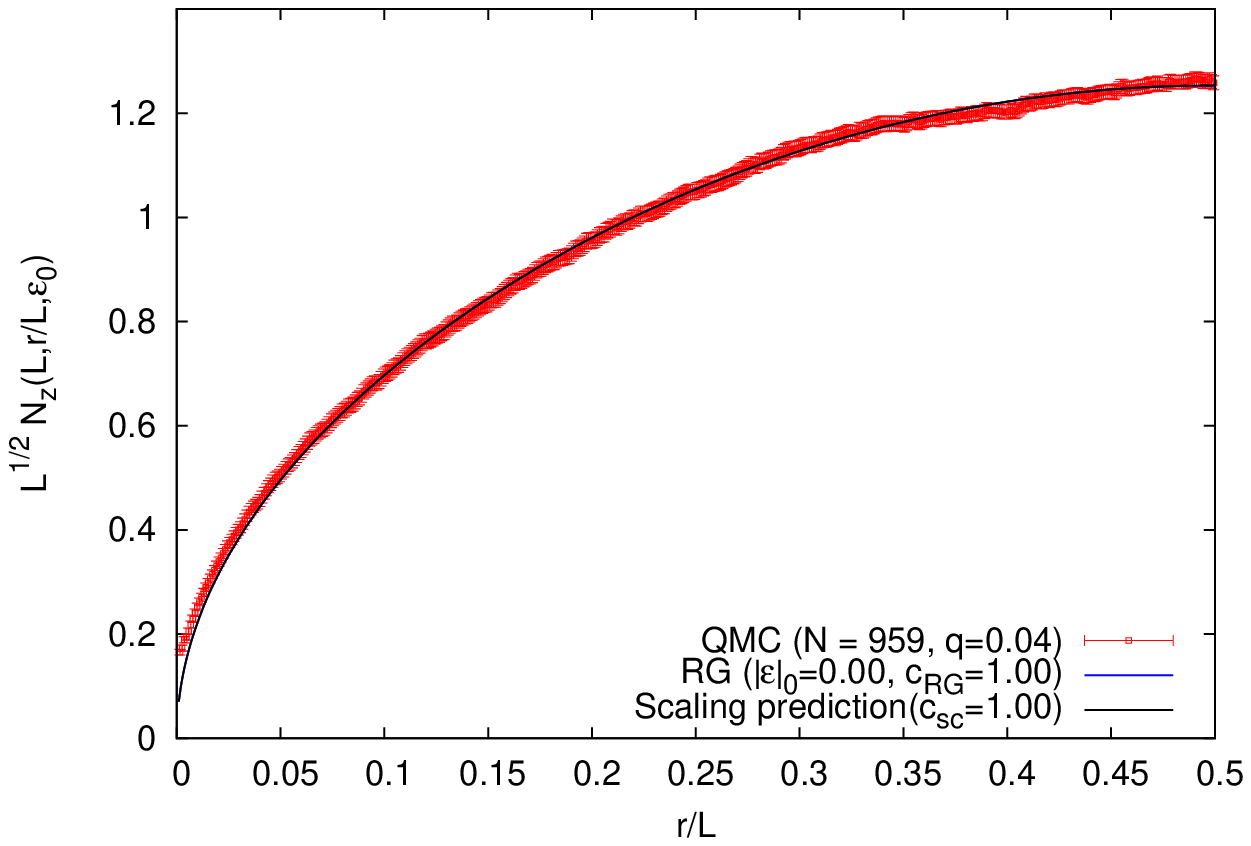}}
\caption{$L^{1/2}N_z(r)$ plotted versus $r/L$ (where $L =N+1$ for
chains with $N=959$ and $N=479$ spins and open boundary conditions) and compared with
the scaling prediction $F_0$ for $q= 0.04$, the approximate
location of the quantum critical point separating the power-law
Neel phase from the VBS ordered phase in the one dimensional
$JQ_3$ model.  Note the data at both sizes fits
essentially perfectly to the scaling
prediction with the same prefactor $c_{sc}$. Also note that
the best two-parameter
fit corresponding to our RG improved perturbation theory result
also gives $|\epsilon_0| = 0$, and thus coincides with the scaling
answer.}
\label{L480960q04}
\end{figure}

The rest of this article is organized as follows: In Section~\ref{analytics},
we first summarize our approach to the
analytical calculation of the ground state spin texture induced by a missing spin,
give our final predictions for the nature of the logarithmic
violations of scaling, and discuss
them from a somewhat more general RG standpoint. In Section~\ref{numerics},
we describe our projector QMC studies and compare the numerical
data for $N_z$
with our analytical predictions to establish our main results. We conclude with a very brief discussion regarding
the connection between our results and earlier work on the
effect of vacancies on the NMR Knight
shift and the spin structure factor.

\section{Bosonization calculation of ground state spin texture}
\label{analytics}
\subsection{Preliminaries}
\label{preliminaries}
As is well-known, we may model our one dimensional magnet by the continuum effective
Hamiltonian\cite{Affleck_review}
\begin{equation}
H=H_0+H_1
\end{equation}
where the free field part $H_0$ is written as
\begin{equation}
 H_{0}=\frac{u}{2}\int^{L}_{0} dx \left[\left(\dfrac{d \phi}{d x}\right)^2+\left(\dfrac{d \widetilde{\phi}}{d x}\right)^2\right]\; ,
\end{equation}
and the interaction term $H_1$ reads
\begin{equation}
H_{1}=-\frac{u \epsilon_0}{r_{0}^{2}}\int^{L}_{0} dx \cos \left(\frac{2 \phi(x)}{R}\right) \; ; 
\end{equation}
here $r_{0}$ is an ultraviolet regulator defined precisely later and
\begin{equation}
 \frac{1}{2 \pi R^{2}}=1-\pi \epsilon_0 \; .
\end{equation}
The last constraint that relates $R$ to the bare
coupling constant $\epsilon_0$ at scale $r_0$ 
arises from the $SU(2)$ spin invariance of the underlying
microscopic theory.\cite{Orignac} The well-known Kosterlitz-Thouless
renormalization group theory\cite{Jose_Kadanoff_Kirkpatrick_Nelson} applied to the
present SU(2) symmetric case yields the flow equation
\begin{equation}
 \dfrac{d \epsilon}{d \log L)}=\beta_{\epsilon}(\epsilon(L))
\end{equation}
with the one loop expression for the beta function being given by\cite{Barzykin_Affleck}
\begin{equation}
 \beta_{\epsilon}(\epsilon(L))=2 \pi \epsilon^{2}(L)-\frac{1}{2}(2 \pi)^{2}) \epsilon^{3}(L)\; .
\end{equation}
This equation can be solved to obtain the running coupling constant
$\epsilon(L)$ at scale $L$ as\cite{Barzykin_Affleck}
\begin{equation}
 \frac{1}{\epsilon(L)}-\frac{1}{\epsilon_0}= - 2 \pi \left\{ \log \left(\frac{L}{r_{0}}\right)+\frac{1}{2}\log \left(\log \left(\frac{L}{r_{0}} \right) \right) \right\}  +O(1)
\end{equation}
Note that $\epsilon_0$ is {\em negative} in the power-law ordered
antiferromagnetic phase in the present sign convention.

Within this bosonized formulation,
the operator $S^z(r)$ at site $r=ja$ is represented as\cite{Furusaki_Hirahara}
\begin{equation}
 S^{z}(r)=\frac{a}{2 \pi R} \frac{d \phi}{d r} + \frac{\mathcal{A}}{\sqrt{r_0}}(-1)^{\frac{r}{a}}\sin \left(\frac{\phi(r)}{R} \right) \; .
\end{equation}
Here, the coefficient of the uniform part is fixed by $SU(2)$ invariance
while  the coefficient of the alternating part is sensitive to
microscopic details: $ \mathcal{A} = \sqrt{a} c$ where $a$ is lattice spacing of lattice model and $c$ is a pure number that depends on the microscopic Hamiltonian.

Finally, we also recall that the 1-point function $S=\langle \frac{1}{\sqrt{r_0}}\sin \left(\frac{\phi(r)}{R} \right)\rangle_{\uparrow}$
of the operator $\frac{1}{\sqrt{r_0}}\sin \left(\frac{\phi(r)}{R}\right)$
can be thought
of as a function of $L$ and the running coupling $\epsilon(L)$ for fixed
bare coupling $\epsilon_0$ and fixed $r/L$.
Thought of in this way, it obeys the
Callan-Symanzik type equation\cite{Barzykin_Affleck}
\begin{equation}
 \left(\frac{\partial}{\partial ln L}+\beta_{\epsilon}(\epsilon)\frac{\partial}{\partial \epsilon}+\gamma(\epsilon)\right)S\left(L, \epsilon(L) \right | \epsilon_0,\frac{r}{L})=0.
\end{equation}
with the anomalous dimension having the expansion
\begin{equation}
 \gamma(\epsilon)=\frac{1}{2}+\left(\frac{\pi}{2} \right)\epsilon(L)
\end{equation}
in terms of the running coupling $\epsilon$.
As is well-known, this can be solved to leading order in $\epsilon(L)$
to give the following scaling law for $S$
\begin{equation}
 S \cong \frac{F_{0}}{\sqrt{L}} \left(\frac{\epsilon_{0}}{\epsilon(L)}\right)^{\frac{1}{4}}(1+\epsilon(L) R) \; ,
\end{equation}
where $F_{0}\left(\frac{r}{L}\right)$ and $R\left(\frac{r}{L}\right)$ are 
some functions of the ratio $\frac{r}{L}$ and the key point about this
formal expression for $S$ is that all dependence on the ultraviolet
regulator $r_0$ has been traded in for a dependence on $\epsilon(L)$,
the running coupling at scale $L$ for a flow that starts with
bare coupling $\epsilon_0$ at scale $r_0$.

\subsection{Overview}
\label{overview}
With these preliminaries out of the way, we now
outline the strategy used below to calculate the alternating part
of $\langle S^z(r)\rangle_{\uparrow}$. The basic idea is to begin
by calculating the result
for this  alternating part using the bosonized part of the alternating
spin density and bare perturbation theory to first order in $\epsilon_0$ for a finite system of length
$L$. As we shall see below,
this bare perturbation
theory result will turn out to depend logarithmically on the value
of the ultraviolet cutoff $r_0$ via a logarithmic ultraviolet divergence
arising from a first order perturbation theory contribution
proportional to $\epsilon_0\log \frac{L}{r_0}$. This logarithmic
divergence makes bare perturbation theory suspect, since a notionally
small ${\mathcal O}(\epsilon_0)$ correction turns out
to have a logarithmically diverging coefficient.

To extract useful information from the bare perturbation theory,
it is therefore necessary to appeal to the Callan-Symanzik equation for
the one point function $S$, and use the fact that $S$ is
expected to have the general form
\begin{equation}
 S \cong \frac{F_{0}}{\sqrt{L}} \left(\frac{\epsilon_{0}}{\epsilon(L)}\right)^{\frac{1}{4}}(1+\epsilon(L) R) \; ,
\end{equation}
as noted earlier. In order to make contact
with our bare perturbation theory result,
we expand this renormalization group prediction to first order in the bare
coupling constant:
\begin{eqnarray}
&&S = \frac{F_{0}\left(\frac{r}{L}\right)}{\sqrt{L}}\left(1-\frac{\pi}{2}\epsilon_{0}\log\frac{L}{r_{0}}+....\right)\left(1+\epsilon_{0} R(\frac{r}{L})+....\right)
\nonumber \\ 
&&\cong \frac{F_{0}(\frac{r}{L})}{\sqrt{L}}\left(1-\frac{\pi}{2}\epsilon_{0}\log\frac{L}{r_{0}}+\epsilon_{0} R(\frac{r}{L})+....\right).
\label{S structure}
\end{eqnarray}
By comparing with the result of our first order perturbation theory
in $\epsilon_0$, it becomes possible to fix the functions $F_0$ and $R$.
This strategy gives us the one-loop RG improved
result for the alternating part of $\langle S^z(r)\rangle_{\uparrow}$
\begin{equation}
N_z(r) = c \sqrt{a}\frac{F_{0}}{\sqrt{L}} \left(\frac{\epsilon_{0}}{\epsilon(L)}\right)^{\frac{1}{4}}(1+\epsilon(L) R) \; ,
\label{prediction}
\end{equation}
with 
\begin{equation}
 F_{0}(\frac{r}{L})=-\sqrt{\frac{\pi \sin \theta_r}{2}} \; ,
\label{prediction1}
\end{equation}
and 
\begin{equation}
R(\frac{r}{L})=\frac{\pi}{2}\log \frac{2\pi }{ \sin \theta_r } + 2 \left( \int_{0}^{\theta_r} + \int_{0}^{\pi-\theta_r} \right)\phi \cot \phi d\phi \; ,
\label{prediction2}
\end{equation}
with $\theta_r \equiv \frac{\pi r}{L}$.

In order to cast this expression into an explicitly useful form
for comparison with numerical results on a chain of $N$ sites
with lattice spacing $a$, we rewrite the prefactor as
\begin{eqnarray}
&& \left(\frac{\epsilon_{0}}{\epsilon(L)}\right)^{\frac{1}{4}}\approx \nonumber \\
&&\left(1+2\pi |\epsilon_{0}|\left\{\log \left(\frac{L}{r_{0}} \right)+\frac{1}{2}\log \left(\log \left(\frac{L}{r_{0}}\right) \right) \right\} \right)^{1/4}\; , \nonumber \\
&&
\end{eqnarray}
express $\epsilon(L)$ as
\begin{equation}
\epsilon(L) = -\frac{|\epsilon_0|}{1+ 2\pi |\epsilon_0|\left\{ \log \left(\frac{L}{r_{0}}\right)+\frac{1}{2}\log \left(\log \left(\frac{L}{r_{0}} \right) \right) \right\} } \; ,
\end{equation}
choose the short-distance cutoff as $r_0=a$, and set the length
$L$ to $L=(N+1)a$ (see subsection~\ref{details} below).
Eqns~(\ref{prediction}),(\ref{prediction1}), (\ref{prediction2})
with these inputs constitutes a theoretical prediction with two free parameters
(the overall amplitude $c$, and the bare coupling $\epsilon_0$
at the lattice scale), and we find below that this provides an extremely
good two-parameter fit of our numerical data in the power-law ordered
antiferromagnetic phase of the one dimensional $JQ_3$ model. In addition,
the spin texture at $q=q_c$, the critical end-point of this power-law
ordered Neel phase, fits extremely well to the scaling function $F_0$,
to which the more general prediction reduces when $\epsilon_0=0$.

What do these results tell us about the possible origins of such
multiplicative logarithmic corrections to spin textures at
other critical points? To explore this, let
us consider the same calculation of the spin texture, but
at a different critical point with an {\em irrelevant}
coupling $g$ with small scaling dimension $\alpha$.
In other words, we assume that $\beta(g) = -\alpha g + \dots$ with
$\alpha$ small and positive,
and $\gamma(g) = \delta_0 + \delta_1 g + \dots$.
In this case, the Callan-Symanzik equation would predict that
$N_z$ satisfy the scaling law
\begin{equation}
N_z(\vec{r}) = \exp\left( -\int_{g_0}^{g(L)} \frac{\gamma(g)}{\beta(g)} dg\right) F\left(\frac{\vec{r}}{L},g(L)\right) 
\end{equation}
for some function $F$ (that needs a more detailed analysis to determine).
Using the postulated form of the $\beta$ and $\gamma$ functions,
one can therefore conclude
\begin{equation}
N_z(\vec{r}) = \frac{{\cal C}}{L^{\delta_0}}F\left(\frac{\vec{r}}{L},g_0/L^{\alpha}\right)
\end{equation}

Thus, if the critical point in question has no marginal operators, the spin texture will quite generally obey scaling
as long as the scaling function $F(x,y)$ does not diverge as $y \rightarrow 0$.
Conversely, if the critical point in question has a marginal operator,
scaling will always be violated by multiplicative logarithmic factors
even if the scaling function $F(x,y)$ is perfectly analytic and
well-defined in the $y \rightarrow 0$ limit. Indeed, in this
marginal case, the only
way of {\em evading} a multiplicative logarithmic
correction would be to ``arrange'' for the $y \rightarrow 0$ limit
of the scaling function $F(x,y)$ to have exactly the ``right''
kind of singularity needed to cancel the effects
of the multiplicative logarithmic correction coming from the exponential
prefactor.  One may therefore conclude that unless the scaling function
has a particularly ``fine-tuned'' form, scaling predictions
for $N_z$ will be {\em generically
violated by multiplicative logarithmic corrections} in the presence
of a marginal operator. Conversely, irrelevant operators can
lead to violations of scaling only if the scaling
function has a divergence as this operator renormalizes to zero.

\subsection{Details}
\label{details}
When a missing-spin defect is introduced into a periodic spin
chain of $N+1$ sites, it converts the system into
a spin chain of $N$ spins obeying open boundary conditions.
These open boundary conditions can be modeled by refering
back to the original periodic system and requiring that the spin density is constrained to go to zero at the missing site. As is well known,\cite{Furusaki_Hirahara,Eggert_Affleck}
this boundary condition can be incorporated by 
expanding the bosonic field $\phi$ in terms of bosonic normal modes
as follows:
\begin{eqnarray}
 \phi(r)&=& \pi R + \frac{q_0}{L} r + \sum_{n=1}^{\infty} \frac{\sin \left(\frac{n \pi r}{L}\right)( a_n+a_{n}^\dagger)}{\sqrt{\pi n}} \\ \nonumber
\widetilde{\phi}(r)&=& \widetilde{\phi}_{0} + i \sum_{n=1}^{\infty}\frac{\cos \left(\frac{n \pi r}{L} \right)(a_n-a_{n}^\dagger)}{\sqrt{\pi n}} \; .
\end{eqnarray}
Here, the non zero bosonic commutation relations are $\left[\widetilde{\phi}(0),q_0 \right]=i$,$\left[a_{m},a_{n}^{\dagger}\right]=\delta_{m n}$, 
$H_0$ can be written (apart from an (infinite) constant $\frac{u}{2} \sum_{n=1}^{\infty} \frac{n \pi}{L}$) in the canonical form
\begin{equation}
 H_{0}=\frac{u}{2} \frac{q_{0}^{2}}{L}+\sum_{n=1}^{\infty}\left(\frac{u n \pi}{L}\right)a_{n}^{\dagger}a_{n} \; .
\end{equation}
Thus the ground state $|G_0\rangle $ of the unperturbed Hamiltonian is the vacuum
for all the $a_n$, and an eigenstate of the zero mode $q_0$.
Indeed, $q_0|G_0\rangle = \pi R |G_0\rangle$ for
the $S_{tot} = 1/2$, $S^z_{tot} =1/2$ ground state that
we wish to model (more generally $|G_0\rangle$ is an
eigenstate of $q_0$ with eigenvalue $2 \pi RS_{tot}^{z}$).

Now, the ground state corrected to first order in $\epsilon_0$ can
be written formally as
\begin{equation}
 |G\rangle \cong |G_0\rangle - \sum_{k \neq G_0} (\frac{\langle k | H_1|G_{0}\rangle}{E_{k}^{0}-E_{G_0}^{0}})|k\rangle.
\end{equation}
Here $k\equiv { \{N_n \}}$ with $n=1,2\dots\infty$ and $N_{n}$ being the
number of bosons in mode $n$.
For an arbitrary excited state, we have the unperturbed energy
\begin{equation}
 E^{0}({\{N_n\}})=\frac{u}{2}\frac{q_{0}^{2}}{L}+\sum_{n}\omega_{n}\left(N_n+\frac{1}{2}\right)
\end{equation}
with $\omega_{n}=\frac{u n \pi}{L}$, which gives us the following
expression for the energy denominators:
\begin{equation}
 E^{0}({\{N_n\}})-E_{g}^{0}=\sum_{n} \omega_{n} N_n.
\end{equation}
As a result, our formal expression for the ground state corrected
to first order in $\epsilon_0$ now reads
\begin{eqnarray}
 &&|G\rangle = \arrowvert {\{N_n=0\}} \rangle +\nonumber \\
&& \frac{u \epsilon_0}{r_0^2}\sum_{\{N_n\}\neq \{0 \}}\left(\frac{\langle{\{N_n\}}\arrowvert \int_{0}^{L} \cos(\frac{2 \phi(x)}{R}) \arrowvert{\{0\}}\rangle}{u\sum_{n} \frac{n \pi}{L} N_n}\right)\arrowvert { \{N_n\}} \rangle \nonumber \\
&&
\label{formalexpression}
\end{eqnarray}

This gives the following formal expression for the one point function:
\begin{widetext}
\begin{eqnarray}
&& S \cong \langle {\{0\}} \arrowvert \frac{1}{\sqrt{r_0}}\sin \left(\frac{ \phi(r)}{R} \right) \arrowvert {\{0\}} \rangle + \frac{\epsilon_0}{r_0^2} \sum_{{\{N_n\}\neq \{0\}}}\frac{\langle {\{0\}} \arrowvert \frac{1}{\sqrt{r_0}}\sin\left(\frac{ \phi(r)}{R}\right) \arrowvert {\{N_n\}} \rangle \langle{\{N_n\}}\arrowvert \int_{0}^{L} dx \cos \left(\frac{2 \phi(x)}{R}\right) \arrowvert{\{0\}}\rangle}{\sum_{n}\frac{n \pi}{L}N_n} \nonumber \\
&&+ \frac{\epsilon_0}{r_0^2} \sum_{\{N_n\}\neq \{0\}}\frac{\langle {\{N_n\}} \arrowvert \frac{1}{\sqrt{r_0}}\sin(\frac{ \phi(r)}{R}) \arrowvert {\{0\}} \rangle \langle{\{0\}}\arrowvert \int_{0}^{L} dx \cos\left(\frac{2 \phi(x)}{R}\right) \arrowvert{\{N_n\}}\rangle}{\sum_{n}\frac{n \pi}{L}N_n} \; , \nonumber \\
&&
\end{eqnarray}
\end{widetext}
where we can set $R=1/\sqrt{2\pi}$ in the contributions that
arise from the ${\mathcal O}(\epsilon_0)$ corrections
to $|G_0\rangle$, as long as we are careful to use
the full expression $R=(2\pi-2\pi^2 \epsilon_0)^{-1/2} \approx (1+ \pi \epsilon_0/2)/\sqrt{2\pi}$ when
evaluating the first ``unperturbed'' term in order to
obtain the latter correct to ${\mathcal O}(\epsilon_0)$.
To evaluate the matrix elements and expectation values, it is
useful to write the state $|\{N_n\}\rangle$ in ``coordinate'' representation as
\begin{equation}
 \langle \{ y_n \} \arrowvert \{ N_n \} \rangle = \prod_{n=1}^{\infty}\left(\frac{1}{\pi^{\frac{1}{4}}2^{\frac{N_n}{2}}}\frac{1}{\sqrt{N_n !}}e^{-\frac{y_n^2}{2}} H_{N_n}(y_n)\right)
\end{equation}
where the coordinates $y_n=\frac{a_{n}+a_{n}^{\dagger}}{\sqrt{2}}$ are
conjugate to ``momenta'' $\pi_n=\frac{a_{n}-a_{n}^{\dagger}}{i \sqrt{2}}$ and
$H_m(x)$ is the $m^{th}$ Hermite polynomial of $x$.
The expectation values in our formal perturbative
expression above can now be evaluated in closed form using
this coordinate representation to obtain the following
compact integral representation of
$S$
\begin{widetext}
\begin{eqnarray}
&&S\left(L,\frac{r}{L},\epsilon_0\right)=-\sqrt{\frac{\pi \sin \theta_r}{2 L }}\left(1-\frac{\pi \epsilon_0}{2 }\log \frac{\pi r_0}{2 L \sin \theta_r }\right) \nonumber \\
&&- \epsilon_0 \sqrt{\frac{\pi \sin \theta_r }{2 L }} \frac{1}{4\sin \theta_r }
\times \left[ \int_{0}^{\infty}\int_{0}^{\pi} d s d\phi\frac{\sin(\theta_r-2\phi)}{\sin^2 \phi } \left( \frac{\cos (\theta_r-\phi)-\cos (\theta_r+\phi)}{\cosh s-\cos (\theta_r-\phi)}  \right) \right] \nonumber \\
&&- \epsilon_0 \sqrt{\frac{\pi \sin \theta_r }{2 L }} \frac{1}{4\sin \theta_r }
\times \left[ \int_{0}^{\infty}\int_{0}^{\pi} ds d\phi \frac{\sin(\theta_r+2\phi)}{\sin^2 \phi}\int_{0}^{\infty} ds \left( \frac{\cos (\theta_r+\phi)-\cos (\theta_r-\phi)}{\cosh s-\cos (\theta_r+\phi)} \right) \right].
\end{eqnarray}
\end{widetext}
Here, $\theta_r \equiv \pi r/L$, and we have regulated mode sums $\sum_{m=1}^{\infty} g_m$ over
the harmonic oscillator modes by replacing them with 
$\sum_{m=1}^{\infty}g_m\exp(-\pi m r_0/L)$ whenever necessary.
It is now possible to do the $s$ integrals in closed form to obtain
the following integral representation for $S$:

\begin{widetext}
\begin{eqnarray}
&& S\left(L,\frac{r}{L},\epsilon_0\right)=-\sqrt{\frac{\pi \sin \theta_r}{2 L }}\left(1-\frac{\pi \epsilon_0}{2 }\log \frac{\pi r_0}{2 L \sin \theta_r }\right) 
+\frac{\epsilon_0}{2}\left( \frac{\pi}{2 L}\right)^{\frac{1}{2}} \int_{0}^{\pi-\theta_r} d\phi \frac{2 \sin \phi \sin \theta_r }{\sqrt{\sin \theta_r }\sin^2 \phi } \sin \left(2\phi+\theta_r\right) \frac{\pi-(\phi+\theta_r)}{\sin \left(\pi-(\phi+\theta_r) \right) }  \nonumber\\
&&\!\!\!\!\!\!\!\!\!\!+ \frac{\epsilon_0}{2}\left( \frac{\pi}{2 L}\right)^{\frac{1}{2}} \left[\int_{\pi-\theta_r}^{\pi} d\phi \frac{2 \sin \phi \sin \theta_r }{\sqrt{\sin \theta_r }\sin^2 \phi } \sin \left(2\phi+\theta_r\right) \frac{(\phi+\theta_r)-\pi}{\sin \left((\phi+\theta_r)-\pi \right) }  + \int_{0}^{\theta_r} d\phi \frac{2 \sin \phi \sin \theta_r }{\sqrt{\sin \theta_r }\sin^2 \phi } \sin \left(2\phi-\theta_r\right) \frac{(\phi-\theta_r)+\pi}{\sin \left((\phi-\theta_r)+\pi \right) } \right]\nonumber \\
&&+ \frac{\epsilon_0}{2}\left( \frac{\pi}{2 L}\right)^{\frac{1}{2}} \int_{\theta_r}^{\pi} d\phi \frac{2 \sin \phi \sin \theta_r }{\sqrt{\sin \theta_r }\sin^2 \phi } \sin \left(2\phi-\theta_r\right) \frac{\pi-(\phi-\theta_r)}{\sin \left(\pi-(\phi-\theta_r) \right) }  \; . \nonumber \\
&&
\end{eqnarray}
\end{widetext}
This integral representation is again regulated with the short
distance cut-off $r_0$ by requiring that the $\phi$ integrals
are to be done by excluding the region $[\theta_r - \pi r_0/L, \theta_r +\pi r_0/L]$ from the integration range. Somewhat remarkably,
it is possible to obtain explicit expressions for all
integrals sensitive to this ultraviolet cutoff, and thereby
reduce this integral representation to the following
compact and simple form:
\begin{widetext}
\begin{eqnarray}
 S\left(L,\frac{r}{L},\epsilon_0\right)&=&-\sqrt{\frac{\pi \sin \theta_r}{2 L }}\left( 1-\frac{\pi \epsilon_0}{2 }\log \frac{L}{r_0}+\frac{\pi \epsilon_0}{2 }\log \frac{2\pi }{ \sin \theta_r } + 2 \epsilon_0 \left( \int_{0}^{\theta_r} + \int_{0}^{\pi-\theta_r} \right)\phi \cot \phi d\phi \right) 
\end{eqnarray}
\end{widetext}
Comparing with the general expectation from our RG analysis (equation \ref{S structure}),  we therefore obtain
\begin{equation}
 F_{0}(\frac{r}{L})=-\sqrt{\frac{\pi \sin \theta_r}{2}}.
\end{equation}
and 
\begin{equation}
R(\frac{r}{L})=\frac{\pi}{2}\log \frac{2\pi }{ \sin \theta_r } + 2 \left( \int_{0}^{\theta_r} + \int_{0}^{\pi-\theta_r} \right)\phi \cot \phi d\phi 
\end{equation}
as already advertised in Section~\ref{overview}.

\section{Numerical computations}
\label{numerics}

Our numerical work on chains with an odd number of sites relies crucially on the spin-half sector generalization~\cite{Banerjee_Damle} of the valence-bond projector QMC algorithm.\cite{Sandvik_Evertz} In our approach, the $S_{tot}=1/2$ sector of the Hilbert
space of an odd number of $S=1/2$ moments, to which the ground
state belongs,  is spanned by a bipartite
valence-bond cover which leaves one spin `free'. Roughly speaking,
the ground state spin texture $\Phi(r) = \langle S^z(r)\rangle_{\uparrow}$ is then obtained directly in our method by keeping track of the probability for the free spin to be at various sites $r$ (see Ref.~\onlinecite{Banerjee_Damle} for details).
\begin{figure}
{\includegraphics[width=\columnwidth]{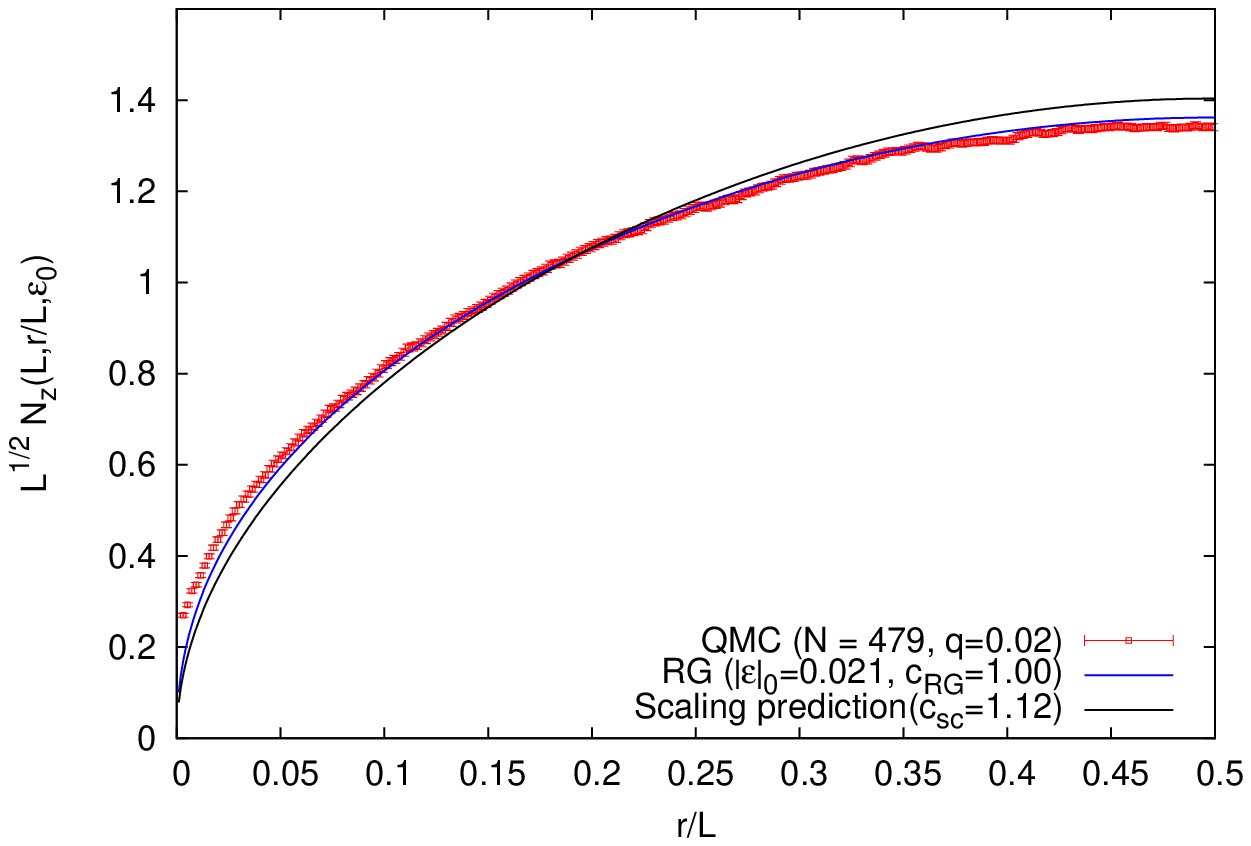}}
{\includegraphics[width=\columnwidth]{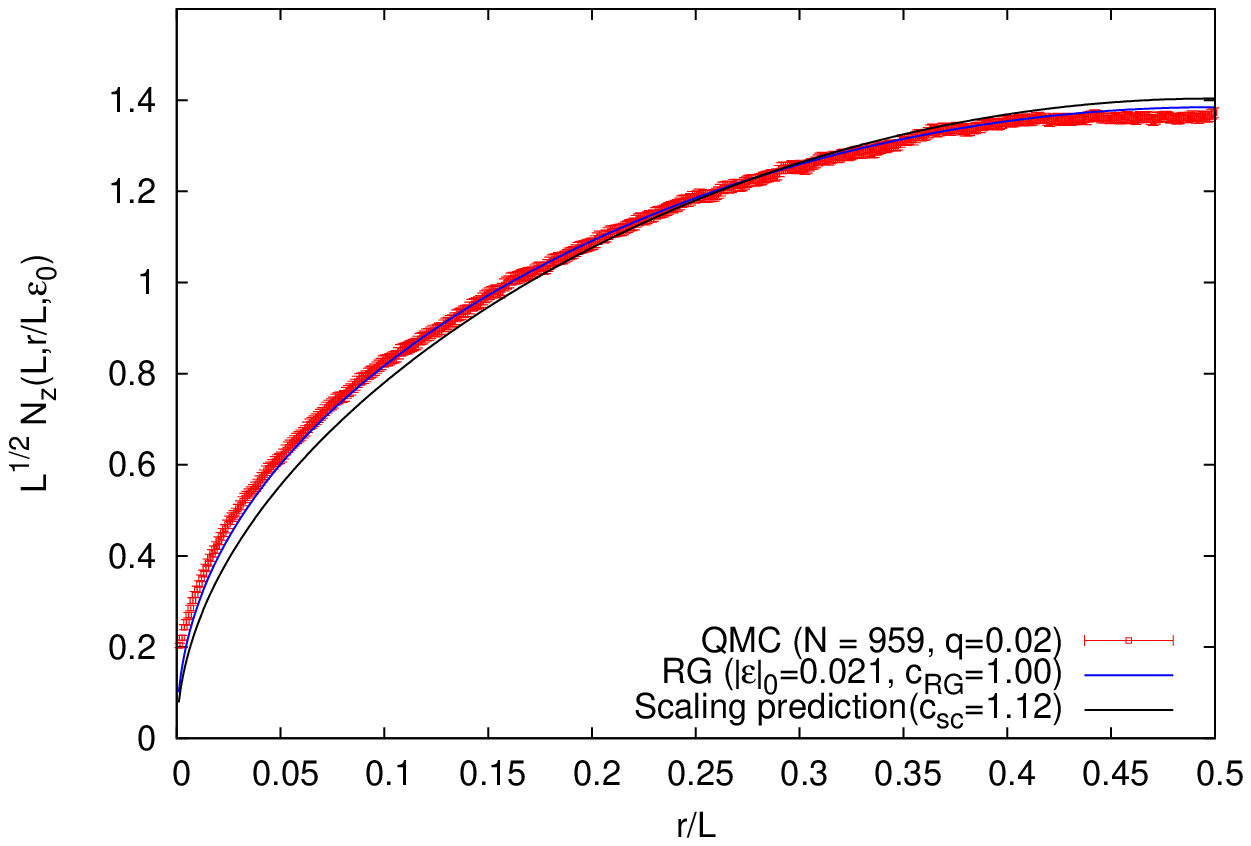}}
\caption{$L^{1/2}N_z(r)$ plotted versus $r/L$ in the power-law ordered
Neel phase at $q= 0.02$ (where $L =N+1$ for
chains with $N=959$ and $N=479$ spins and open boundary conditions) and compared with
the scaling prediction with a common best fit prefactor $c_{sc}$. Note that the deviation of the data from the scaling
prediction cannot be simply ascribed to an overall multiplicative
factor that grows with $N$, since the {\em shape} of the curves
is slightly different. Data at both sizes is also fit to the best two-parameter
fit corresponding to our RG improved perturbation theory result, and
the agreement is seen to be excellent for the best fit
values of $c_{RG}$ and $|\epsilon_0|$ listed in the legend.}
\label{L480960q02}
\end{figure}

This method has also been used in computations of ground state
spin textures at `deconfined' critical points in two dimensional
$SU(2)$ and $SU(3)$ antiferromagnets\cite{Banerjee_Damle_Alet1,Banerjee_Damle_Alet2}, as well as in very recent parallel work on developing
a diagnostic for the presence of sharply-defined spinon
excitations\cite{Ying_Sandvik} in antiferromagnets.

We consider pure systems with periodic boundary conditions and total number of sites ranging from $N=60$ to $N=960$, as well as the corresponding
open spin chains obtained by removing one site from the pure system.
Our projection power is
chosen to scale as $4N^3$ to ensure convergence to the ground state. We perform  $\gtrsim10^5$ equilibration 
steps followed by $\gtrsim10^6$ Monte Carlo measurements to ensure that
statistical errors are under control. In systems
with a vacancy, we measure $\Phi(r)$ in the manner outlined above,
and coarse-grain over pairs
of successive sites to obtain our numerical
results for the alternating part $N_z(r)$, which is to be
thought of as living on bond-centers in this coarse-graining procedure. 
We have checked that
our conclusions are not sensitive to the precise coarse-graining procedure
used, although non-universal details, such as the overall amplitude
of $N_z(r)$, do change.
\begin{figure}
{\includegraphics[width=\columnwidth]{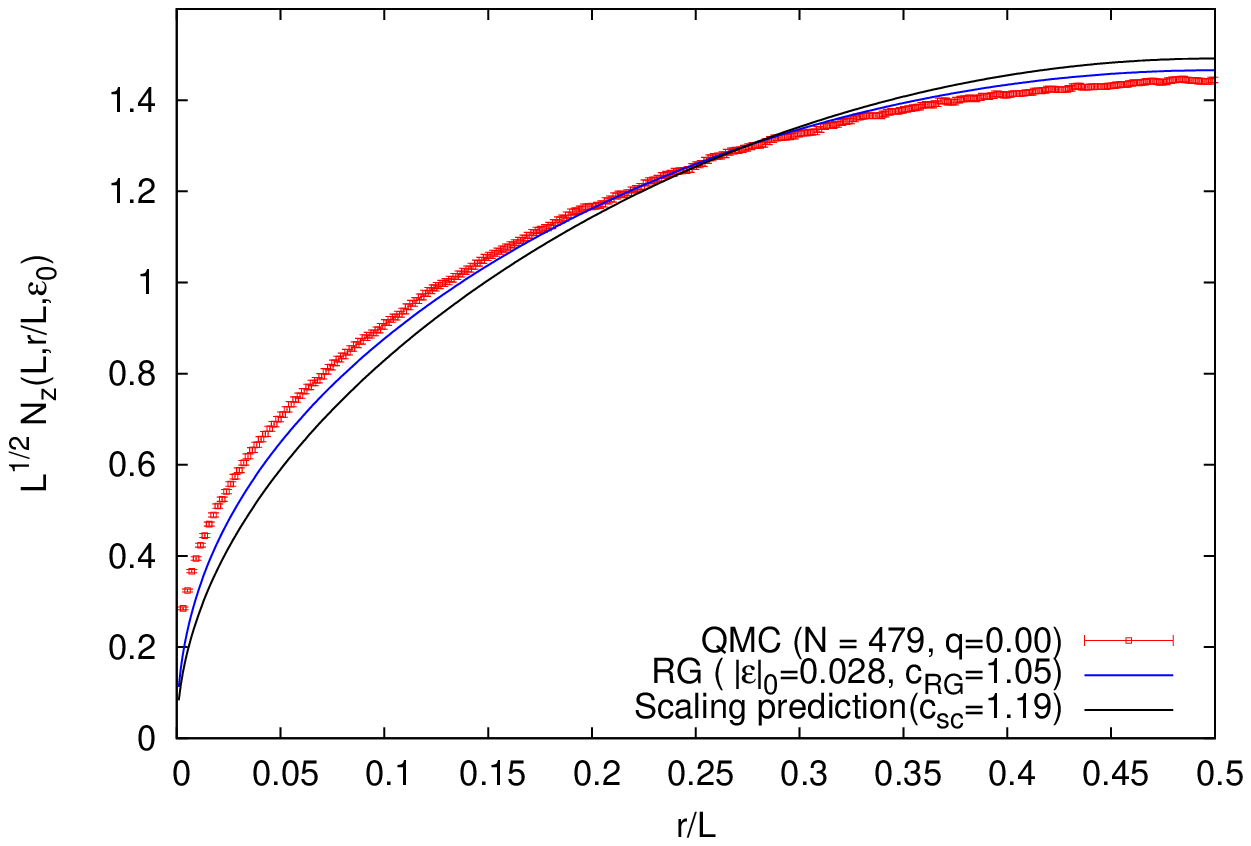}}
{\includegraphics[width=\columnwidth]{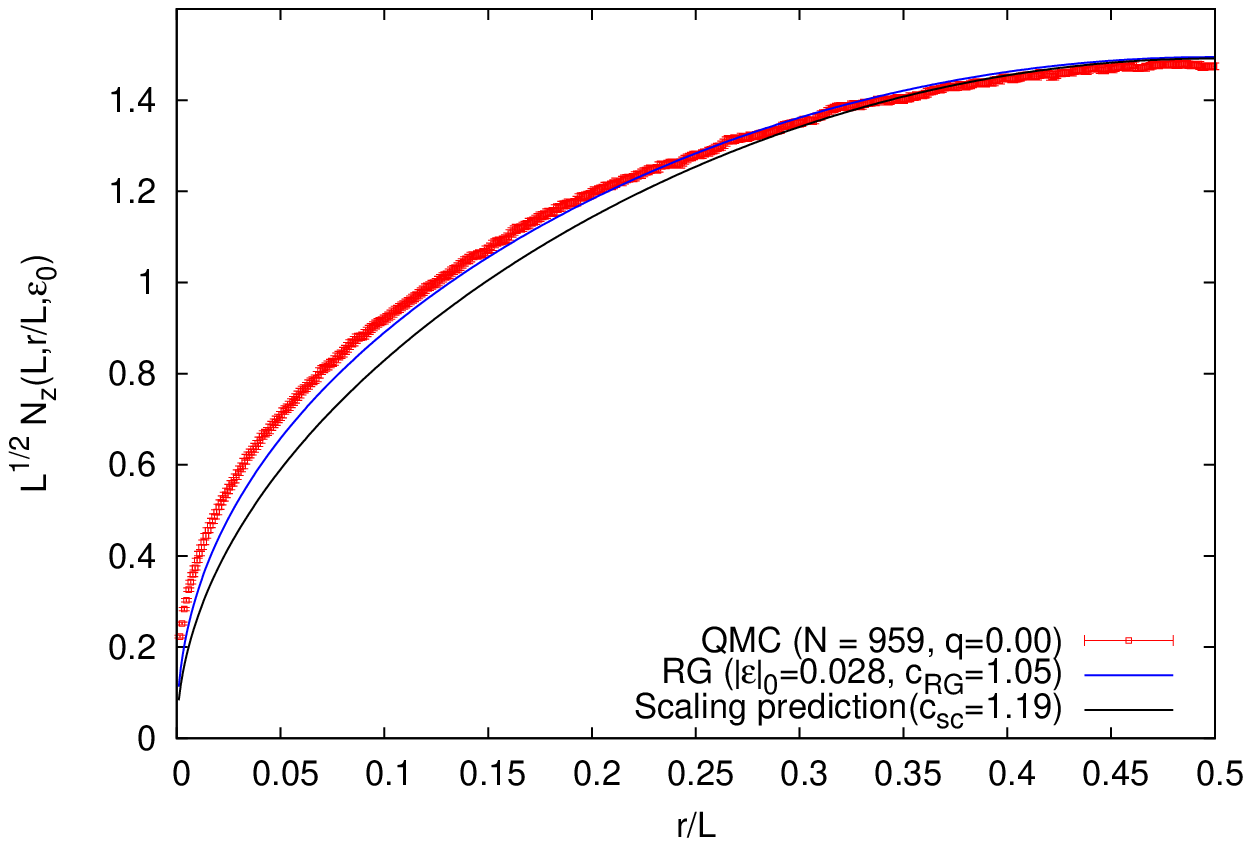}}
\caption{$L^{1/2}N_z(r)$ plotted versus $r/L$ in the power-law ordered
Neel phase at $q= 0.0$ (where $L =N+1$ for
chains with $N=959$ and $N=479$ spins and open boundary conditions) and compared with
the scaling prediction with a common best-fit prefactor $c_{sc}$. Note that the deviation of the data from the scaling
prediction cannot be simply ascribed to an overall multiplicative
factor that grows with $N$, since the {\em shape} of the curves
is slightly different. Data at both sizes is also fit to the best two-parameter
fit corresponding to our RG improved perturbation theory result, and
the agreement is seen to be quite reasonable, but not perfect, for the best fit
values of $c_{RG}$ and $|\epsilon_0|$ listed in the legend.}
\label{L480960q00}
\end{figure}

For the corresponding pure systems, we employ the singlet sector
valence bond projection QMC technique\cite{Sandvik_Evertz}, and
calculate the ground state spin-spin correlation function $C(j)= \langle \vec{S}(0) \cdot \vec{S}(j)\rangle$ for two
sites separated by $j-1$ intervening sites ($j \leq N/2$, where $N$ is the total number
of spins).
To begin with, we scan the six-spin coupling $q=Q/4J$ and study
the $q$ and $N$ dependence of $NC(N/2)$ as a convenient
diagnostic that distinguishes the power-law Neel ordered phase
at small $q$ from the spontaneously dimerized valence bond solid (VBS) ordered phase that is stabilized for large $q$.
In the power-law Neel phase, $NC(N/2)$ grows (logarithmically)
slowly with $N$, while in the VBS phase, it fall off
rapidly with increasing $N$.
Precisely at the critical point separating these two phases, we thus
expect a crossing point for $NC(N/2)$ plotted against $q$ for
various values of $N$.
This is precisely what is seen in our data shown in Fig~\ref{crossing}.
From our data, we estimate that the critical point separating
these two phases is located at $q_c \approx 0.04$ with
an error of approximately $0.005$ estimated by extrapolating
for the position of the crossing point (this estimate
is consistent with the critical point found in Ref.~\onlinecite{Ying_Sandvik}).

With this in hand, we compute the ground state spin texture in
the corresponding chains with one site removed
for several $q \leq  q_c$ for a range of system sizes.
The alternating part of the computed spin texture
is then compared with the scaling predictions obtained
by setting $\epsilon_0 =0$,  as well as with
our RG improved perturbation theory predictions. The former
represents a one-parameter fit of the data, with the overall
amplitude $c$ being the only free parameter, while
the latter should be thought of as a two parameter fit,
with the bare value $\epsilon_0$ of the sine-Gordon coupling
being the second fitting parameter. 

In Fig~(\ref{L480960q04}), we first display
our data for the alternating part of the spin texture
and compare it with the scaling prediction at the putative critical
point $q=q_c$ for two of our largest system sizes. As is
clear from these two figures, the scaling prediction fits
extremely well to all the data at both sizes. Furthermore,
a two-parameter fit using the RG-improved perturbation
theory result yields a best-fit value of $\epsilon_0$
indistinguishable from $\epsilon_0 = 0$. This confirms
our identification of the critical point, since
we expect that the bare coefficient of the marginally irrelevant
cosine interaction is zero at this quantum phase transition.

This excellent fit to the scaling prediction should
be contrasted with the results shown in
Figs~(\ref{L480960q02}),(\ref{L480960q00}),
which show numerical results at two representative points in the power-law Neel phase compared with the one-parameter fit obtained from
the scaling prediction. As is clear from these
figures, the scaling prediction simply cannot provide a satisfactory
account of the data for $q < q_c$, with the discrepancy being more pronounced
for smaller $q$, that is,  {\em further away} from the critical point.
Furthermore, the observed deviations from scaling cannot
be simply ascribed to an overall $N$ dependent prefactor that grows with system
size, since the {\em shapes} of the curves are themselves slightly different
from the scaling prediction.

In the same figures,
we also show the best two-parameter fit obtained by using our
RG improved perturbation theory result. Two points are worth
noting regarding these two parameter fits: Firstly, the best-fit values
of $|\epsilon_0|$ increase as one goes further away from $q=q_c$,
consistent with the expectation that the bare coefficient
of the cosine interaction vanishes as $q$ approaches $q_c$.
Second, the RG improved perturbation theory provides
a much better fit at $q=0.02$ than at the Heisenberg point $q=0$---again
this is consistent with our expectations, since our calculation
is perturbative in the renormalized coupling $\epsilon(L)$,
and is therefore expected to provide a better approximation when the bare
value of $|\epsilon_0|$ is smaller to begin with.

\section{Discussion}
\label{discussion}
We conclude by clarifying the relationship of our calculations
with earlier calculations of the effect of vacancies on \cite{Sirker_Laflorencie,Eggert_Affleck_impurities,Brunel_Bocquet_Jolicoeur,Fujimoto_Eggert}
 on spin chains. These have typically focused
on the low-field NMR Knight shift and relaxation rate $1/T_1$ in the presence of vacancies, or the impurity contribution to the zero-field spin structure factor and linear
susceptibility of such chains. All these experimental observables are
obtained from the knowledge of the zero field static and equal time spin correlations of the system at finite temperature, which has been the main focus of
this body of work. In contrast, our results focus
on local spin texture induced by the presence of vacancies
{\em at } $T=0$, which is a quite different observable connected
with the impurity contribution to the local susceptibility in the high-field
regime in which the external field dominates over the thermal fluctuations.

\section{Acknowledgements}
We acknowledge useful discussions with Ribhu Kaul, Nicolas Laflorencie, Gautam Mandal,
Anders Sandvik, and Diptiman Sen, computational resources of the TIFR,
and support from DST (India) grant DST-SR/S2/RJN-25/2006.


\begin{thebibliography}{999}

\bibitem{exptchains} M.~Takigawa, N.~Motoyama, H.~Eisaki, and S.~Uchida,
Phys. Rev. B {\bf 55}, 14129 (1997).

\bibitem{exptchains1} D.~A.~Tennant {\em et. al.}, Phys. Rev. B {\bf 71}, 134412 (2005). 

\bibitem{exptchains2} B.~Lake, D.~A.~Tennant, C.~D.~Frost, S.~E.~Nagler,
Nature Materials {\bf 4}, 329 (2005). 

\bibitem{exptchains3} B.~Lake, D.~A.~Tennant, S.~E.~Nagler,
Phys. Rev. B {\bf 71}, 134412 (2005).
 

\bibitem{Affleck_review} I.~Affleck, {\em Fields, Strings, and Critical Phenomena}, Les Houches 1988, E.~Brezin and J.~Zinn-Justin (eds.), North-Holland, Amsterdam (1990).

\bibitem{Singh_Fisher_Shankar} R.~R.~P.~Singh, M.~E.~Fisher, and
R.~Shankar, Phys. Rev. B {\bf 39}, 2562 (1989).

\bibitem{Affleck_Gepner_Shultz_Ziman} I.~Affleck, D.~Gepner, H.~Shultz,
and T.~Ziman, J Phys. A: Math. Gen. {\bf 22}, 511 (1989).

\bibitem{Barzykin_Affleck} V.~Barzykin and I.~Affleck, J Phys. A: Math. Gen. {\bf 32}, 867 (1999).


\bibitem{Orignac} E.~Orignac, Eur. Phys. J. B {\bf 39}, 335 (2004).

\bibitem{Diptiman} R. Chitra, S.~Pati,
H.~R.~Krishnamurthy, D.~Sen, and S.~Ramasesha, Phys. Rev. B {\bf 52}, 6581 (1995). 

\bibitem{Eggert} S.~Eggert, Phys. Rev. B {\bf 54}, R9612 (1996).

\bibitem{Kenna_ONmodel} R.~Kenna, Nuclear Physics B {\bf 691 [FS]}, 292 (2004). 

\bibitem{Sachdev} S.~Sachdev, Phys. Rev. B {\bf 50}, 13006 (1994).



\bibitem{Sandvik} A.~W.~Sandvik, Phys. Rev. Lett. {\bf 104}, 177201 (2010).

\bibitem{Banerjee_Damle_Alet1} A.~Banerjee, K.~Damle, and F.~Alet
Phys. Rev. B {\bf 82}, 155139 (2010).

\bibitem{Senthil_etal} T. Senthil, L.~Balents, S.~Sachdev, A.~Vishwanath, and M.~P.~A.~Fisher, Phys. Rev. B {\bf 70}, 144407 (2004).

\bibitem{Sandvik_original} A.~W.~Sandvik, Phys. Rev. Lett. {\bf 98}, 227202 (2007).


\bibitem{Kaul} R.~K.~Kaul, arXiv:1010.1937, unpublished.

\bibitem{Sandvik_Kotov_Sushkov} A.~W.~Sandvik, V.~N.~Kotov, and O.~P.~Sushkov,
Phys. Rev. Lett. {\bf 106}, 207203 (2011).

\bibitem{Banerjee_Damle_Alet2} A.~Banerjee, K.~Damle, and F.~Alet,
Phys. Rev. B {\bf 83}, 235111 (2011).

\bibitem{Sandvik_SU3} J.~Lou, A.~W.~Sandvik, N.~Kawashima,
Phys. Rev. B {\bf 80}, 180414(R) (2009).


\bibitem{Sandvik_Evertz} A.~W.~Sandvik, and  H.~G.~Evertz, Phys. Rev. B {\bf 82}, 024407 (2010).

\bibitem{Banerjee_Damle} A. Banerjee and K. Damle, J. Stat. Mech. (2010) P08017.

\bibitem{Jose_Kadanoff_Kirkpatrick_Nelson} J.~V.~José, L.~P.~Kadanoff, S.~Kirkpatrick, and D.~R.~Nelson, Phys. Rev. B {\bf 16}, 1217 (1977).

\bibitem{Furusaki_Hirahara} T.~Hirahara and A.~Furusaki, Phys. Rev. B {\bf 63},
134438 (2001).

\bibitem{Eggert_Affleck} S.~Eggert and I.~Affleck, Phys. Rev. B {\bf 46},
10866 (1992).


\bibitem{Ying_Sandvik} Y.~Tang and A.~W.~Sandvik, unpublished.




\bibitem{Sirker_Laflorencie}J.~Sirker and N.~Laflorencie, Europhys. Lett. {\bf 86}, 57004 (2009).

\bibitem{Eggert_Affleck_impurities} S.~Eggert and I.~Affleck, Phys. Rev.
Lett. {\bf 75}, 934 (1995).

\bibitem{Brunel_Bocquet_Jolicoeur} V.~Brunel, M.~Bocquet, and Th.~Jolicoeur,
Phys. Rev. Lett. {\bf 83}, 2821 (1999).

\bibitem{Fujimoto_Eggert} S.~Fujimoto and S.~Eggert, Phys. Rev. Lett. {\bf 92},
037206 (2004).










\end{thebibliography}
\end{document}